\documentclass[11pt,tightenlines,aps,prd,groupedaddress,nofootinbib,nobibnotes,longbibliography,notitlepage]{revtex4-1}
\pdfoutput=1

\usepackage{graphicx}
\usepackage{amsfonts}
\usepackage{amssymb}
\usepackage{amsbsy}
\usepackage{amsmath}
\usepackage{mathtools}
\usepackage{latexsym}
\usepackage{bm}
\usepackage{color}
\usepackage{hyperref}
\usepackage[margin=2cm]{geometry}
\usepackage{comment}
\usepackage[makeroom]{cancel}
\usepackage[dvipsnames]{xcolor}

\usepackage{stackengine}
\stackMath
\newcommand\tenq[2][1]{%
 \def\useanchorwidth{T}%
  \ifnum#1>1%
    \stackon[0pt]{\tenq[\numexpr#1-1\relax]{#2}}{\scriptscriptstyle\sim}%
  \else%
    \stackon[1pt]{#2}{\scriptscriptstyle\sim}%
  \fi%
}

\newcommand{\de}{\mbox{d}}
\newcommand{\lf}{\left}
\newcommand{\rg}{\right}
\newcommand{\be}{\begin{equation}}
\newcommand{\ee}{\end{equation}}
\newcommand{\pha}{\phantom{a}}

\newcommand{\pa}{\partial}

\newcommand{\bea}{\begin{eqnarray}}
\newcommand{\eea}{\end{eqnarray}}

\newcommand{\arctanh}{{\rm arctanh}}

\numberwithin{equation}{section}

\begin{document}

\title{\Large Cosmological evolution from modified Bekenstein entropy law}

\author{Marco de Cesare}
\email{marco.decesare@na.infn.it}

\author{Giulia Gubitosi}
\email{giulia.gubitosi@unina.it}

\affiliation{Dipartimento di Fisica ``Ettore Pancini'', Universit{\`a} di Napoli ``Federico II'', Napoli, Italy}
\affiliation{INFN, Sezione di Napoli, Italy}


\begin{abstract}
We study the dynamics of the homogeneous and isotropic cosmological background in the recently proposed ``quantum phenomenological gravitational dynamics'', characterised by logarithmic corrections to the Bekenstein entropy.
We show that the model admits a family of solutions that are self-accelerating both at early and late times: they approach de Sitter in the future and admit a past attractor corresponding to an inflationary acceleration era. On the other hand, there are no solutions corresponding to a primordial bounce. We also show that asking scalar perturbations to be unaffected by instabilities on observable scales puts stringent constraints on the deviations from general relativity encoded by the model.

\end{abstract}

\maketitle

\tableofcontents

\nopagebreak

\newpage

\section{Introduction}

Effective modified gravitational dynamics can be linked to theories where the Bekenstein entropy law acquires quantum corrections, a feature shared by several approaches to quantum gravity~\cite{Kaul:2000kf,Carlip:2000nv,Meissner:2004ju,Sen:2012dw}. A notable case is that of Loop Quantum Gravity: on the one hand, it predicts a modified Bekenstein law~\cite{Kaul:2000kf,Meissner:2004ju}; on the other hand, the modified Friedmann equations that describe the background cosmological evolution in its cosmological realization~\cite{Ashtekar:2011ni,Ashtekar:2015iza}
replace the primordial singularity with a bounce. Therefore, a natural question that arises is whether a cosmological bounce is a generic consequence of theories with a modified Bekenstein law.

Recent work \cite{Alonso-Serrano:2020dcz} derived a ``quantum phenomenological gravitational dynamics" within the approach known as thermodynamics of spacetime \cite{Jacobson:1995ab}, including a correction term in the Bekenstein entropy law that is logarithmic term in the area, of the same kind as the one found in Loop Quantum Gravity~\cite{Kaul:2000kf,Meissner:2004ju}.
This approach leads to a modification of the field equations of general relativity that include higher-order curvature invariants, leading to correction terms in the Friedmann equations that are governed by a free parameter. In this work, we investigate whether such modified gravitational dynamics predict a primordial bounce.
In fact, it was suggested in Ref.~\cite{Alonso-Serrano:2020dcz} and more recently in Ref.~\cite{Alonso-Serrano:2022nmb} that in this model the initial cosmological singularity may be replaced by a regular bounce.
However, such conclusions are based on a perturbative analysis of the dynamics of the cosmological background, where departures from general relativity are assumed to be small~\cite{Alonso-Serrano:2020dcz}. In this work we show that such an assumption does not hold at all times, and in particular it fails in the early universe.

In this work, we use dynamical system techniques to analyze the full space of solutions of the exact field equations for the cosmological background, described as a spatially flat Friedmann-Lema{\^i}tre-Robertson-Walker (FLRW) spacetime filled with hydrodynamical matter. 
Our analysis shows that the system does not admit any non-singular bouncing solution for the cosmological background. It is however interesting that the system admits a family of solutions that are self-accelerating both at early and late times.

Our requirements to identify cosmologically viable solutions, without introducing exotic matter that violates the energy conditions, are the following:
(i) the universe undergoes an epoch of accelerated expansion at late times and (ii) departures from general relativity be suppressed during earlier epochs of matter and radiation domination, so as not to spoil the predictions of the $\Lambda$CDM model. Moreover, if the model is to offer an alternative to the standard inflationary scenario in the early universe (i.e., before reheating), we shall also require that (iii) the horizon problem is addressed.

Condition (iii) can be met either by means of an inflationary epoch that lasts for a sufficiently large number of e-folds, or by means of a non-singular bounce (whereby the horizon problem is automatically solved \cite{Brandenberger:2016vhg}). In the case of inflationary solutions we require that the ``slow-roll conditions'' be satisfied ($\epsilon\equiv-\dot{H}/H^2\ll1$, $\eta\equiv\dot{\epsilon}/(H\epsilon)\ll1$) and that the number of e-folds be at least of order $N\approx 60$. In the case of bouncing solutions, the universe makes a smooth transition from a contracting phase (where the Hubble rate is negative, $H<0$) to an expanding phase ($H>0$); this implies that the Hubble rate $H$ vanishes at a finite time $t_{\rm\scriptscriptstyle B}$, where it has a positive time derivative $\dot{H}(t_{\rm\scriptscriptstyle B})>0$.

Concerning the solutions we find to be viable, they approach de Sitter in the future and admit a past attractor corresponding to an inflationary era.
For this class of solutions, slow-roll conditions can be satisfied in the early universe depending on the matter content and the value of the free parameter of the model. However, the slow-roll conditions can only be satisfied if the equation-of-state parameter of matter is extremely close to $w=-1$. For this reason, unless we assume exotic matter in the early universe (such as a scalar field), the model is {\it per se} unable to provide a satisfactory alternative to the inflationary scenario---already at the background level.

In order to investigate cosmological viability at late times (i.e., during radiation domination and subsequent epochs), we also analyze the dynamics of cosmological perturbations to check for the presence of instabilities, focusing on scalar and tensor perturbations.
We show that the equations for scalar perturbations are modified compared to standard cosmology and that such modifications are responsible for instabilities. By imposing the requirement that observable scales are not affected by such instabilities, we obtain a bound on the free parameter of the model.
We also derive the equations of motion for tensor perturbations, showing that they have a non-standard coupling to anisotropic stress. 

Lastly, we note that the field equations proposed in Ref.~\cite{Alonso-Serrano:2020dcz} are traceless, and can therefore be seen as a generalization of the trace-free Einstein equations~\cite{Ellis:2010uc}.\footnote{The trace-free Einstein equations are also closely linked to unimodular gravity \cite{HENNEAUX1989195,Kuchar:1991aa,Percacci:2017fsy}.} In analogy with the trace-free Einstein equations, also in this model the cosmological constant arises as an integration constant. Moreover, the model here considered admits solutions that are accelerating both at late times and at early times---without the need to introduce exotic matter to sustain the accelerated expansion. This is an interesting feature of the model that is not shared by trace-free Einstein gravity, where an inflationary epoch without scalar fields \cite{Leon:2022kwn} can only be achieved at the cost of violating energy-momentum conservation. Note though that in trace-free Einstein gravity (and in unimodular gravity), energy-momentum conservation is not a consequence of the field equations; rather, it must be imposed as an independent assumption \cite{Ellis:2010uc,Josset:2016vrq}. This is also true in the model at hand, due to the tracelessness of the field equations. 
Within the context of trace-free Einstein gravity and unimodular gravity, this allows for a natural embedding of a special class of interacting dark energy models (so-called interacting vacuum) which has been studied in Refs.~\cite{Perez:2020cwa,deCesare:2021wmk}. From the point of view of an effective description of quantum gravity based on unimodular gravity, it has been argued in Ref.~\cite{Perez:2017krv} that the violation of energy-momentum conservation that sources dark energy may be ascribed to a fundamentally discrete spacetime structure.
In the present paper we follow Refs.~\cite{Alonso-Serrano:2020dcz,Alonso-Serrano:2022nmb} and assume that the energy-momentum of matter is strictly conserved, although generalizations are possible.

The plan of the paper is as follows. In Section~\ref{Sec:Review} we review the formulation of ``quantum phenomenological gravitational dynamics'' proposed in Ref.~\cite{Alonso-Serrano:2020dcz}. The evolution of the cosmological background is studied in detail in Section~\ref{Sec:Background}. The equations for scalar and tensor perturbations are derived and analyzed in Sections~\ref{Sec:Scalar}, \ref{Sec:Tensor}, respectively. Lastly, our results are reviewed in Section~\ref{Sec:Discussion}. 

\

{\bf Conventions} We use the metric signature $(-+++)$. We choose units such that $c=\hbar=1$. The gravitational coupling is $\kappa=8\pi G$. An overdot is used to denote derivatives w.r.t.~proper time, whereas a prime is used for derivatives w.r.t.~conformal time.

\section{Dynamical equations of ``quantum phenomenological gravitational dynamics''}\label{Sec:Review}
The field equations proposed in Ref.~\cite{Alonso-Serrano:2020dcz} read as
\be\label{Eq:FieldEquations}
S_{\mu\nu} - \alpha \kappa S_{\mu\lambda}S^\lambda_{\pha \nu}+\frac{\alpha \kappa}{4}\lf( R_{\kappa \lambda}R^{\kappa \lambda} -\frac{1}{4}R^2\rg) g_{\mu\nu}=\kappa \lf( T_{\mu\nu}-\frac{1}{4}T g_{\mu\nu}\rg)~,
\ee
where $S_{\mu\nu}=R_{\mu\nu} - (R/4) g_{\mu\nu}$ is the traceless part of the Ricci tensor, $T_{\mu\nu}$ is the stress-energy tensor of matter, and $\alpha$ is a dimensionless coupling.\footnote{In Refs.~\cite{Alonso-Serrano:2020dcz,Alonso-Serrano:2022nmb}, the couplings corresponding to the higher-order curvature terms in \eqref{Eq:FieldEquations} is $D \ell_{\rm P}^2$, where $\ell_{\rm P}$ is the Planck length and $D$ is dimensionless. Thus, with our choice of units we have $\alpha=D/(8\pi)$.} The numerical coefficient $\alpha$ is given by the pre-factor of the logarithmic correction to the Bekenstein entropy \cite{Alonso-Serrano:2020dcz},  and is treated as a free parameter to be determined phenomenologically.

The field equations~\eqref{Eq:FieldEquations} are a generalization of the trace-free Einstein equations \cite{Ellis:2010uc}, which are recovered for $\alpha=0$. In both theories (and in contrast with general relativity) the continuity equation for matter fields does not follow from the field equations via the Bianchi identities; rather, it must be imposed by hand as an independent assumption.\footnote{Nonetheless, this condition can be relaxed to allow for dynamical dark energy with $w=-1$, see Refs.~\cite{Josset:2016vrq,Perez:2020cwa,deCesare:2021wmk}.} The cosmological constant then arises as an integration constant, as will be clarified in Section~\ref{Sec:Background}.

We assume that matter can be described as a perfect fluid, with stress-energy tensor
\be\label{Eq:StressEnergy}
T_{\mu\nu}=(\rho+p)u_{\mu}u_{\nu}+p\, g_{\mu\nu}~,
\ee
where $u_{\mu}$ is the four-velocity (normalized as $u_\mu u^\mu=-1$), $\rho$ is the energy density, and $p$ is the pressure. Only in Section~\ref{Sec:Tensor} we will introduce an anisotropic stress component to source tensor perturbations.
As in Ref.~\cite{Alonso-Serrano:2020dcz}, we also assume that $T_{\mu\nu}$ is covariantly conserved, although we stress that---unlike general relativity---this is an independent assumption that does not follow from the field equations \eqref{Eq:FieldEquations}
\be\label{Eq:Continuity}
\nabla^\mu T_{\mu\nu}=0~.
\ee

\section{Evolution of the cosmological background}\label{Sec:Background}
We model the cosmological background as a FLRW spacetime with zero spatial curvature, described by the line element
\be\label{Eq:FLRWmetric}
\de s^2= - \de t^2+  a(t)^2 \delta_{ij} \de x^i\de x^j ~,
\ee
where $a(t)$ is the scale factor, $x^i$ are comoving coordinates and $t$ represents proper time. (In the following, the time dependence will be omitted to make the notation lighter.)
With the line element \eqref{Eq:FLRWmetric}, the field equations~\eqref{Eq:FieldEquations} boil down to just one independent equation
\be\label{Eq:Hprime}
\dot{H}(1-\alpha \kappa \dot{H})=-\frac{\kappa}{2}(\bar{\rho}+\bar{p})~,
\ee
where $H=\dot{a}/a$ is the Hubble rate and an overbar is used to denote background fluid quantities. 
Equation~\eqref{Eq:Hprime} represents a generalization of the Raychaudhuri equation.
We note that in Ref.~\cite{Alonso-Serrano:2020dcz,Alonso-Serrano:2022nmb}, a modified Friedmann equation was derived from Eq.~\eqref{Eq:Hprime} within a perturbative approach, i.e.~treating deviations from general relativity as small correction terms. Instead, for our analysis of the background dynamics we focus exclusively on Eq.~\eqref{Eq:Hprime}, which is derived from the field equations~\eqref{Eq:FieldEquations} without any approximations.

Furthermore, Eq.~\eqref{Eq:Continuity} gives
\be\label{Eq:Rhoprime}
\dot{\bar{\rho}}+3 H (\bar{\rho}+\bar{p})=0~.
\ee
Thus, the background cosmological dynamics is given by the autonomous dynamical system \eqref{Eq:Hprime}, \eqref{Eq:Rhoprime}. In the following, we will assume for simplicity that matter has a barotropic equation of state, $p=w\rho$ with constant $w$, subject to the null energy condition $w+1>0$.

We need to distinguish two cases, based on the sign of the parameter $\alpha$. These will be analysed in detail in the remainder of this section.

\subsection{Case I: $\alpha>0$}

In order to recast the system \eqref{Eq:Hprime}, \eqref{Eq:Rhoprime} in a form amenable to be studied using dynamical system techniques, 
we start by observing that equation~\eqref{Eq:Hprime} can be regarded as a constraint on $\dot{H}$, $\bar{\rho}$, which suggests the following parametrization
\begin{align}
\dot{H}&=\frac{1}{2\alpha \kappa}\lf(1\pm \cosh\varphi\rg)~,\label{Eq:ParametrizationHubble}\\
\bar{\rho}&=\frac{1}{2\alpha \kappa^2(1+w)}(\sinh\varphi)^2~,\label{Eq:ParametrizationRho}
\end{align}
with $\varphi\geq0$.\footnote{Given the symmetry of Eqs.~\eqref{Eq:ParametrizationHubble}, \eqref{Eq:ParametrizationRho} under the transformations $\varphi\to-\varphi$, it would be redundant to extend the domain of $\varphi$ to the whole real axis.}
Substituting the expression for $\bar{\rho}$ given in Eq.~\eqref{Eq:ParametrizationRho} into the continuity equation~\eqref{Eq:Rhoprime} we obtain the equation for $\varphi$
\be\label{Eq:PhiEvolution}
\dot{\varphi}=-\frac{3}{2}(1+w)H \tanh\varphi~.
\ee
Equations~\eqref{Eq:ParametrizationHubble}, \eqref{Eq:PhiEvolution} constitute an autonomous dynamical system for the variables $\varphi$ and $H$, written in normal form.

We observe that the ``plus'' branch in Eq.~\eqref{Eq:ParametrizationHubble} does not give rise to viable solutions, since it implies $\dot{H}>0$ at all times; this branch is also incompatible with general relativity in the limit $\alpha\to0$, since the standard Raychuadhuri equation is not recovered.
Hence, in the following we will only focus on the ``minus branch'', whereby $\dot{H}\leq0$ at all times. Clearly, the latter does not admit bouncing solutions, since these would require that $H$ vanishes at a point where $\dot{H}>0$.

The equation for the orbits can be obtained combining Eqs.~\eqref{Eq:ParametrizationHubble},~\eqref{Eq:PhiEvolution}
\be\label{Eq:Orbits}
\frac{\de H^2}{\de \varphi}=-\frac{2}{3\alpha\kappa (w+1)}\frac{1-\cosh\varphi}{\tanh \varphi}~.
\ee
Integrating, we have
\be\label{Eq:SolOrbits}
\frac{3}{2}(1+w)\alpha \kappa(H^2-c)+1-\cosh\varphi+2\log\cosh\lf(\frac{\varphi}{2}\rg)=0~,
\ee
where $c$ is an integration constant. For positive values of $c$, the orbits intersect the $H$ axis at a non-zero value given by $H_0=\sqrt{c}$. On the other hand, for negative values of $c$, the orbits never reach the $H$ axis; however, they interesect the $\varphi$ axis at a non-zero value of $\varphi$. Lastly, for $c=0$, we obtain a separatrix between the two regimes, intersecting both axes at the origin. We also note that, for a given value of $c$, the orbit only depends on the combination $(1+w)\alpha$.

Before we proceed further, it is convenient to introduce new variables with compact range
\be
U=\tanh\varphi~,\quad V=\tanh(\sqrt{\kappa}H) ~.
\ee
In this way, the configuration space is mapped into the rectangle $0 \leq U \leq 1$ and $-1\leq V \leq 1$. We also introduce a dimensionless time variable $T=t/\sqrt{\kappa}$. Thus, we can recast the dynamical system \eqref{Eq:ParametrizationHubble}, \eqref{Eq:PhiEvolution} in the following form
\begin{subequations}\label{Eq:DynSys1}
\begin{align}
\frac{\de U}{\de T}&=-\frac{3}{2}(1+w)U(1-U^2)\arctanh V~,\\
\frac{\de V}{\de T}&=\frac{1}{2\alpha}\lf(\frac{1}{\sqrt{1-U^2}}-1\rg)(1-V^2)~.
\end{align}
\end{subequations}

\subsubsection{Phase portrait, fixed points and attractors}\label{Sec:PhasePortrait1}

\begin{figure}
\includegraphics[width=.25\columnwidth]{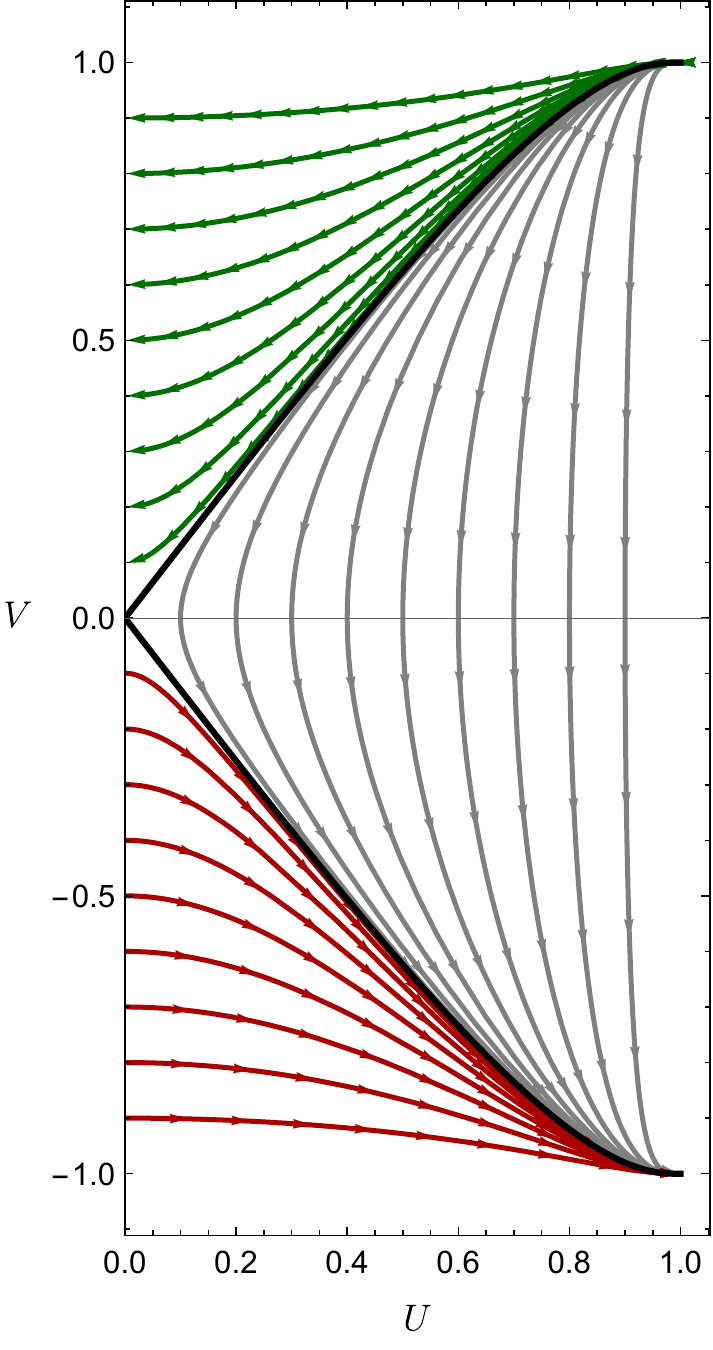}
\caption{Phase portrait of the system with $\alpha>0$. The plot is obtained for $\alpha=10^{-1}$ and $w=0$, but the qualitative features remain the same for generic values of the parameters. Also note that Eq.~\eqref{Eq:Orbits} implies that there is a degeneracy between $\alpha$ and $w$.}
\label{Fig:PhasePortrait_PosAlpha}
\end{figure}

The phase portrait of the system \eqref{Eq:DynSys1} for the $\alpha>0$ case is shown in Figure~\ref{Fig:PhasePortrait_PosAlpha}.
The fixed points of the system are non-isolated and all lie on the segment $U=0$ with $-1\leq V\leq 1$: this is the {\it equilibrium set} of the system \cite{wainwright_ellis_1997}. These points correspond to a universe dominated by a cosmological constant and are only approached asymptotically by the system. In the proximity of the fixed points we have
\be\label{Eq:NearFixedPoints}
V\approx V_0~,\quad U\sim \exp\lf(-\frac{3}{2}(1+w)(\arctanh V_0) T\rg)~.
\ee
Therefore, fixed points with $V_0>0$ can be approached in the $T\to+\infty$ limit, whereas fixed points with $V_0<0$ can be approached in the $T\to-\infty$ limit; that is, the upper and lower portions of the $V$ axis are a {\it local sink} and a {\it local source}, respectively. 
In terms of $H$ and $\bar{\rho}$, Eqs.~\eqref{Eq:NearFixedPoints} read
\be
H\approx H_0=\frac{1}{\sqrt{\kappa}}\arctanh V_0 ~,\quad \rho\sim \exp\lf(-3(1+w) H_0 t\rg)~.
\ee
Thus, there is an effective cosmological constant $\Lambda_{\rm eff}=3H_0^2$ that arises as an integration constant and dominates the large-$T$ dynamics for these orbits. As pointed out in Section~\ref{Sec:Review}, this is to be expected based on the tracelessness of the field equations.

The system has two local attractors, that asymptotically approach the top-right and bottom-right corners of the rectangle in Fig.~\ref{Fig:PhasePortrait_PosAlpha} (respectively, a past and a future attractor). Their equations can be obtained from Eqs.~\eqref{Eq:Orbits} and \eqref{Eq:PhiEvolution} in the large $\varphi$ limit
\be\label{Eq:DynSys1attractor1}
\frac{\de H^2}{\de \varphi}\approx \frac{1}{3\alpha\kappa (w+1)}e^{\varphi}~,\quad
\dot{\varphi}\approx -\frac{3}{2}(1+w)H ~,
\ee
whose solutions are
\begin{subequations}\label{Eq:DynSys1attractor2}
\begin{align}
H\approx \frac{4}{3(1+w)(t-\bar{t}\,)}~,\quad \varphi\approx -2 \log\lf(\frac{1}{4}\sqrt{\frac{3(1+w)}{\alpha\kappa}}|t-\bar{t}|\rg)~.
\end{align}
\end{subequations}
The asymptotics of the energy density in this regime is therefore $\bar{\rho}\sim (t-\bar{t})^{-4}$. In the case of the past attractor, the limit $t\to\bar{t}$ is approached from the right, whereas for the future attractor the limit $t\to\bar{t}$ is approached from the left. Hence, the past attractor corresponds to an expanding universe, whereas the future attractor corresponds to a contracting one. In both cases, there is a curvature singularity: the Ricci scalar has the asymptotics
\be
R=6(\dot{H}+2H^2)\approx \frac{8(5-3w)}{3(1+w)^2}\frac{1}{(t-\bar{t}\,)^2}~.
\ee
Since all orbits in the system approach at least one attractor (either in the past or in the future), this completes the proof that the model does not allow for singularity resolution.

Orbits lying to the right of the separatrix (depicted in gray in Fig.~\ref{Fig:PhasePortrait_PosAlpha}) are excluded, since they do not recover standard cosmology at late times, i.e., the corrections to general relativity parametrized by $\alpha$ are never sub-dominant. These orbits describe a recollapsing universe, eventually approaching the future attractor ending with a Big Crunch singularity. 
It is also not possible to recover an intermediate stage of dark energy domination\footnote{That is, an extended time interval where $H$ is approximately constant.} before recollapse.
Similarly, orbits below the separatrix (in red in Fig.~\ref{Fig:PhasePortrait_PosAlpha}) describe an ever-contracting universe approaching the future attractor; hence, they are also excluded.

This leaves us with the orbits lying above the separatrix (in green in Fig.~\ref{Fig:PhasePortrait_PosAlpha}), that describe an expanding universe that is accelerating at late times, where it approaches de Sitter, and meets the past attractor at early times.
We observe that on the past attractor the evolution of the scale factor obtained from Eq.~\eqref{Eq:DynSys1attractor2}  is the same as in power-law inflation if matter obeys a non-relativistic equation of state, i.e.~we have $a(t)\sim (t-\bar{t}\,)^q$, with $q=\frac{4}{3(1+w)}$, which implies that $\ddot{a}/a>0$ if $w<1/3$. On the other hand, if the universe is radiation dominated, then $w=1/3$ and one has $\ddot{a}=0$, while for larger values of $w$ the universe is decelerating also in this early stage. Recalling the definition of the slow-roll parameter $\epsilon$, we have $\epsilon\equiv-\dot{H}/H^2=1/q$. Hence, in our case the slow-roll condition $\epsilon\ll1$ is satisfied if the dominant matter species in the early universe has $0<1+w\ll1$. 
However, in the absence of scalar fields or other matter species that may possess a negative pressure, this last family of solutions should also be considered as ruled out.

This early epoch of accelerated expansion ends when the correction term to the modified Raychaudhuri equation \eqref{Eq:Hprime} stop beings dominant, i.e.~when $\alpha \kappa |\dot{H}|\approx 1$. This happens at $t_{\rm end}-\bar{t}\approx 2 \sqrt{\alpha \kappa/(3(w+1))}= \sqrt{32\pi\alpha /(3(w+1))}\,t_{\rm Pl} $~. The number of e-folds of inflation obtained {since the end of the Planck era} is
\be
N= \int_{\bar{t}+t_{\rm Pl}}^{t_{\rm end}} \de t\, H = \frac{2}{3(w+1)} \log\lf(\frac{32\pi\alpha}{3(w+1)}\rg)~.
\ee
Thus, in order to obtain $N\approx60$ e-folds of inflation we need to have $\alpha\approx 3(1+w) e^{90(1+w)} /(32\pi)\approx 3(1+w) /(32\pi)$, where in the last step we used $1+w\ll1$, as implied by the slow-roll condition. This implies that, in order to solve the horizon problem in this scenario, not only we need to assume matter with an equation of state $0<1+w\ll1$ in the early universe, but we also need to tune the value of $\alpha$ to produce the correct number of e-folds.

\subsection{Case II: $\alpha<0$}

When $\alpha<0$ the form of Eq.~\eqref{Eq:Hprime} suggests a different parametrization
\begin{align}
\dot{H}&=\frac{1}{2|\alpha| \kappa}\lf(\cos\varphi-1\rg)~,\label{Eq:ParametrizationHubble2}\\
\bar{\rho}&=\frac{1}{2|\alpha| \kappa^2(1+w)}(\sin\varphi)^2~.\label{Eq:ParametrizationRho2}
\end{align}
Combining Eqs.~\eqref{Eq:ParametrizationRho2} and \eqref{Eq:Rhoprime} we obtain
\be\label{Eq:Phiprime2}
\dot{\varphi}=-\frac{3}{2}(1+w)H \tan\varphi~.
\ee
Hence, taking into account the symmetries of the r.h.s.~of Eqs.~\eqref{Eq:ParametrizationHubble2} and \eqref{Eq:Phiprime2}, $\varphi$ can be restricted to the domain $0<\varphi<\pi$.
We note that Eq.~\eqref{Eq:ParametrizationHubble2} implies that $\dot{H}\leq0$ at all times. Therefore, also in this case a cosmological bounce is not possible.

The orbits satisfy the equation
\be
\frac{\de H^2}{\de \varphi}=\frac{2}{3|\alpha| \kappa(1+w)}\frac{1-\cos\varphi}{\tan\varphi}~,
\ee
whose solution is
\be\label{Eq:Separatrix_NegAlpha}
\frac{3}{2}(1+w)|\alpha| \kappa H^2+\cos\varphi-2\log\cos\lf(\frac{\varphi}{2}\rg)=c~,
\ee
where $c$ is an integration constant. The $\varphi<\pi/2$ portion of the curve \eqref{Eq:Separatrix_NegAlpha} with $c=1$ corresponds to the separatrix shown in Fig.~\ref{Fig:PhasePortrait_NegAlpha}. Also in this case the orbits described by Eq.~\eqref{Eq:Separatrix_NegAlpha}, for any given value of $c$, only depend on the combination $(1+w)|\alpha|$.

The system \eqref{Eq:ParametrizationHubble2}, \eqref{Eq:Phiprime2} can be recast in dimensionless form by introducing the dimensionless compact variable $W=\tanh\lf(\sqrt{\kappa}H\rg)$ and dimensionless time $T=t/\sqrt{\kappa}$
\begin{subequations}\label{Eq:DynSys2}
\begin{align}
\frac{\de W}{\de T}&=\frac{1}{2|\alpha|}\lf(\cos\varphi-1\rg)(1-W^2)~,\\
\frac{\de\varphi}{\de T}&=-\frac{3}{2}(1+w)\arctanh W \tan\varphi~.
\end{align}
\end{subequations}

\subsubsection{Phase portrait and fixed points}
\begin{figure}
\includegraphics[width=.5\columnwidth]{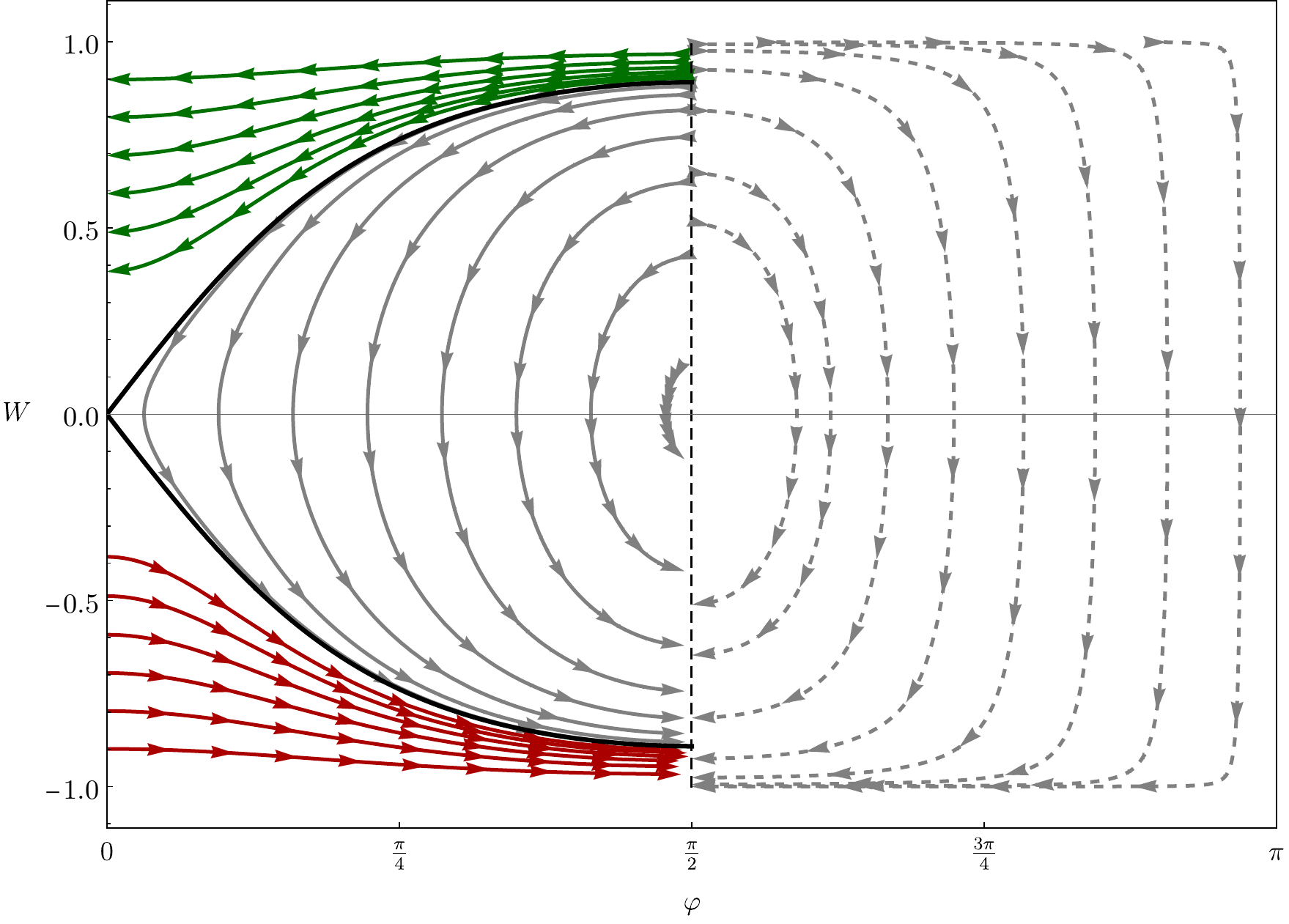}
\caption{Phase portrait of the system with $\alpha<0$. The plot is obtained for $\alpha=-10^{-1}$ and $w=0$, but the qualitative features are general.}
\label{Fig:PhasePortrait_NegAlpha}
\end{figure}

The phase portrait of the system with $\alpha<0$ is shown in Fig.~\ref{Fig:PhasePortrait_NegAlpha}. The dynamical system \eqref{Eq:DynSys2} has non-isolated fixed points, corresponding to the $\varphi=0$ segment with $-1\leq W\leq 1$. Asymptotically, for $\varphi\approx0$ the solutions are
\be
W\approx W_0~,\quad \varphi \sim \exp\lf(-\frac{3}{2}(1+w)\arctanh W_0\, T\rg)~.
\ee
The energy density goes as
\be
\bar{\rho}\sim \exp\lf(-3(1+w)\arctanh W_0\, T\rg)~.
\ee
Thus, the fixed points correspond to a universe dominated by a cosmological constant, with constant Hubble rate $H_0=\arctanh W_0/\sqrt{\kappa}$~. The equilibrium set $\varphi=0$ is approached in the distant future $T\to+\infty$ if $W_0>0$ (local sink), or in the distant past $T\to-\infty$ if $W_0<0$ (local source).

Orbits cannot cross the singular line $\varphi=\pi/2$, where the r.h.s.~of Eq.~\eqref{Eq:Phiprime2} diverges.
As we approach this line, the energy density \eqref{Eq:ParametrizationRho2} attains a maximum
\be\label{Eq:MaximalRho}
\rho_{\rm max}=\frac{1}{2|\alpha| \kappa^2(1+w)}~.
\ee
In the proximity of $\varphi=\pi/2$, solutions have the leading-order asymptotics
\be\label{Eq:AsymptoticsWphi}
\arctanh W\approx\arctanh W_0-\frac{(T-T_0)}{2|\alpha|}~, \quad \lf(\frac{\pi}{2}-\varphi\rg)^2\approx 3(1+w)\arctanh W_0 (T-T_0)~.
\ee
The Ricci scalar is also finite in this limit
\be
R\approx \frac{3}{\kappa} \lf(4\,\arctanh W_0^2 -\frac{1}{|\alpha|}\rg)~.
\ee
Nonetheless, the geometry is still singular in the limit; in fact, from Eqs.~\eqref{Eq:ParametrizationHubble2} and \eqref{Eq:Phiprime2} we obtain 
\be
\ddot{H}= \frac{3(1+w) }{4|\alpha|\kappa}\frac{H \sin^2\varphi}{\cos\varphi}~,
\ee
which is divergent as $\varphi$ tends to $\pi/2$.

Solutions with $0<\varphi<\pi/2$ lying inside the separatrix, as well as solutions with $\pi/2<\varphi<\pi$ correspond to a universe that is initially quasi-de Sitter with positive Hubble rate, but then recollapses and reaches a final quasi-de Sitter state with negative Hubble rate; as such, they are excluded. Furthermore, solutions with $0<\varphi<\pi/2$ and $W<0$ lying below the separatrix are also unphysical, as they describe a universe where the Hubble rate is negative at all times.

\subsubsection{Early-time accelerated expansion}
This leaves as a viable option only the family of solutions with $0<\varphi<\pi/2$ lying above the separatrix.
These solutions start (at a finite initial time) with positive Hubble rate $H_{\rm i}>H_{\rm crit}$, where the critical value is $H_{\rm crit}=\sqrt{\frac{2(1-\log2)}{3|\alpha|\kappa(1+w)}}\sim \frac{M_{\rm Pl}}{|\alpha|(1+w)}$. Correspondingly, the maximal energy density is given by Eq.~\eqref{Eq:MaximalRho}. After this initial stage of acceleration, the solutions then decelerate and eventually reach a final quasi-de Sitter state with positive $H_{\rm f}<H_{\rm i}$. This evolution is qualitatively similar to dynamical dark energy models.
It is therefore natural to investigate whether the early-time background dynamics obtained in this model is comparable to the predictions of inflation.

From Eqs.~\eqref{Eq:ParametrizationHubble2}, \eqref{Eq:Phiprime2} we obtain the following early-time evolution for the Hubble rate and $\varphi$ on the orbits of interest
\begin{subequations}
\begin{align}
H&\approx H_{\rm i}-\frac{t-t_0}{2|\alpha| \kappa}+\sqrt{\frac{(1+w)H_{\rm i}}{3}}\,\frac{(t-t_0)^{3/2}}{|\alpha| \kappa}-\frac{1}{20}\bigg[\frac{1}{2\alpha^2\kappa^2}\sqrt{\frac{3(w+1)}{H_{\rm i}}}+\frac{1}{|\alpha|\kappa}\lf(3(w+1)H_{\rm i}\rg)^{3/2}  \bigg](t-t_0)^{5/2}~,\label{Eq:HubbleEarlyAsymptotics} \\
\varphi&\approx \frac{\pi}{2}-\sqrt{3(1+w)H_{\rm i}}\,(t-t_0)^{1/2}+\frac{1}{4}\sqrt{3(1+w)H_{\rm i}}\lf(\frac{1}{2|\alpha|\kappa H_{\rm i}}+H_{\rm i}(1+w)\rg)(t-t_0)^{3/2}~.
\end{align}
\end{subequations}
Where $t_0$ is the initial (proper) time. 
The slow-roll parameters obtained from Eq.~\eqref{Eq:HubbleEarlyAsymptotics} are
\begin{subequations}
\begin{align}
\epsilon&=-\dot{H}/H^2= \frac{1}{2|\alpha|\kappa H_{\rm i}^2}\lf(1-\sqrt{3(1+w)H_{\rm i}}(t-t_0)^{1/2} +\frac{(t-t_0)}{|\alpha| \kappa H_{\rm i}}\rg)+\mathcal{O}(t-t_0)^{3/2}~,\label{Eq:EpsilonAsymptotics}\\
\eta&=\frac{\dot{\epsilon}}{H\epsilon}= -\frac{1}{2}\sqrt{\frac{3(1+w)}{H_{i}}}\,(t-t_0)^{-1/2}+\frac{1}{|\alpha|\kappa H_{\rm i}^2}-\frac{3}{2}(1+w)+\mathcal{O}(t-t_0)^{1/2}~.\label{Eq:EtaAsymptotics}
\end{align}
\end{subequations}
We observe that it is not possible to satisfy \emph{both} slow-roll conditions $\epsilon\ll1$, $|\eta|\ll1$. In particular, $\eta$ is divergent in the $t\to t_0$ limit, {except for the case where $w=-1$ exactly}. {For values $w\neq-1$}, even if we restricted our attention only to times such that the first term in Eq.~\eqref{Eq:EtaAsymptotics} be negligible, and constrained the value of $\alpha$ so that the overall factor in \eqref{Eq:EpsilonAsymptotics} be much less than unity, we would find that $\eta$ is $\mathcal{O}(1)$, {unless one also assumes $w+1\ll 1$}.
Hence, the early-time background evolution predicted by this family of solutions cannot solve the horizon problem and is therefore ruled out as a possible alternative to the standard inflationary scenario\footnote{These solutions would only be viable when $w\approx-1$, which therefore requires an embedding within the standard inflationary scenario.}
Therefore, also the model with $\alpha<0$ does not provide a viable description of early universe cosmology.

\subsection{Summary}
We showed that in both cases $\alpha\lessgtr0$ the model is unable to offer cosmologically viable solutions for the background at early times {unless exotic matter is assumed}. Nonetheless, we can identify in both cases families of solutions that are potentially viable in the regime where the corrections parametrized by $\alpha$ become sub-leading compared to ordinary matter. These solutions are represented by the green curves in Figs.~\ref{Fig:PhasePortrait_PosAlpha} and \ref{Fig:PhasePortrait_NegAlpha} .

In order to have an evolution for the cosmological background that is consistent with standard cosmology at late times, i.e.~when ordinary matter becomes dominant, we shall require that the $\alpha$-dependent correction terms in Eq.~\eqref{Eq:Hprime} be sub-leading. Thus, we must have $|\alpha\kappa \dot{H}|\ll 1$. Moreover, since we have $\dot{H}<0$ at all times, it is sufficient to impose this constraint at the time of reheating, at the beginning of the radiation dominated era. Thus, we obtain\footnote{Recall that during radiation domination we have $\dot{H}=-2H^2$.}
\be\label{Eq:AlphaConstraint}
|\alpha|\ll \frac{1}{2\kappa H_{\rm reh}^2}=\frac{M_{\rm Pl}^2}{2H_{\rm reh}^2}~,
\ee
where $H_{\rm reh}$ is the Hubble rate at reheating and $M_{\rm Pl}=1/{\sqrt{\kappa}}$ is the Planck mass.

\section{Scalar Perturbations}\label{Sec:Scalar}

Having identified background solutions that are potentially cosmologically viable (after reheating), we now turn to the analysis of cosmological perturbations. Theoretical constraints can be derived from the requirement that perturbations are well behaved.
To this aim, we study the dynamics of scalar perturbations in the longitudinal gauge, using conformal time for the background geometry
\be\label{Eq:FLRWmetricScalarPert}
\de s^2= a(\eta)^2 \Big(- (1+2\phi(\eta,x))\de \eta^2+ (1-2\psi(\eta,x)) \delta_{ij} \de x^i\de x^j \Big) ~,
\ee
The conformal Hubble rate is defined as $\mathcal{H}=a^{\prime}/a$, and we use a prime to denote derivatives w.r.t.~$\eta$~; a comma denotes derivatives w.r.t.~comoving spatial coordinates. In conformal time, the background equation \eqref{Eq:Hprime} reads as
\be
{\cal H}^{\prime}-{\cal H}^2-\frac{\alpha \kappa}{a^2}\lf({\cal H}^{\prime}-{\cal H}^2\rg)^2=-\frac{\kappa}{2}a^2\lf(\bar{\rho}+\bar{p}\rg)~.
\ee
The perturbed stress-energy tensor of matter is
\be
\delta T^0_{\pha 0}=-\delta \rho~,\quad \delta T^0_{\pha i}=\lf(\bar{\rho}+\bar{p}\rg)v_{,i}~,\quad\ \delta T^i_{\pha j}=\delta p\, \delta^i_{\; j}~,
\ee
where $v$ is the velocity potential. For the sake of simplicity, we assume that the scalar part of the anisotropic stress be zero (this is a reasonable approximation for all matter fields except neutrinos). The pressure perturbation can be decomposed as $\delta p=c_s^2\delta\rho+(\delta p)_{\rm nad}$~, where $c_s^2$ is the sound speed and $(\delta p)_{\rm nad}$ is the non-adiabatic pressure perturbation.

From Eq.~\eqref{Eq:FieldEquations} we obtain the following independent equations for perturbations
\begin{subequations}\label{Eq:ScalarPerturbations}
\begin{align}
&\psi-\phi=0~,\\
&\lf[1+\frac{\alpha\kappa}{a^2} \lf({\cal H}^2 -{\cal H}^\prime\rg)\rg] \lf(\phi^{\prime}+{\cal H}\phi\rg)=-\frac{\kappa}{2}a^2\lf(\bar{\rho}+\bar{p}\rg)v ~,\\
&\lf[1+\frac{2\alpha\kappa}{a^2}\lf({\cal H}^2-{\cal H}^{\prime}\rg)\rg]\lf(\phi^{\prime\prime}+2({\cal H}^{\prime}-{\cal H}^2)\phi^{\prime}+\Delta\phi\rg)=\frac{\kappa}{2}a^2\lf(\delta\rho+\delta p\rg)~,
\end{align}
\end{subequations}
where $\Delta=\delta_{ij}\pa_i\pa_j$ denotes the spatial Laplacian. From the matter continuity equation~\eqref{Eq:Continuity} we obtain
\begin{subequations}\label{Eq:FluidPerturbations}
\begin{align}
\lf(\bar{\rho}+\bar{p}\rg)\lf(v^{\prime}+4{\cal H}v+\phi\rg)+\lf(\bar{\rho}+\bar{p}\rg)^{\prime}v+\delta p=0~,\\
\delta\rho^{\prime}-3\lf(\bar{\rho}+\bar{p}\rg)\phi^{\prime}+3{\cal H}(\delta\rho+\delta p)+\lf(\bar{\rho}+\bar{p}\rg)\Delta v=0~.
\end{align}
\end{subequations}

Combining the above equations we obtain the following second order equation for the gravitational potential $\phi$
\be\label{Eq:SecondOrderPDEPhi}
\begin{split}
&\Big(1+\frac{\alpha \kappa}{a^2}(1+c_s^2)({\cal H}^2-{\cal H}^{\prime})\Big)\phi^{\prime\prime}+\Big(3(1+c_s^2){\cal H}+\frac{\alpha \kappa}{a^2}(1+c_s^2) \lf({\cal H}^3+{\cal H}{\cal H}^{\prime}-{\cal H}^{\prime\prime}\rg)  \Big)\phi^{\prime}+\\
&\qquad\Big( (1+3c_s^2){\cal H}^2+2 {\cal H}^{\prime}-\frac{\alpha \kappa}{a^2}(1+c_s^2)\lf({\cal H}^4-5{\cal H}^2{\cal H}^{\prime}+2({\cal H}^{\prime})^2+{\cal H}{\cal H}^{\prime\prime}\rg)\Big)\phi-c_s^2\Delta\phi=\frac{\kappa}{2}a^2(\delta p)_{\rm nad}~.
\end{split}
\ee

\subsection{Evolution of the gravitational potential during the radiation dominated era}\label{Eq:ScalarPert_RDE}
During radiation domination we have $a\sim \eta$, ${\cal H}=1/\eta$, if we impose the constraint \eqref{Eq:AlphaConstraint} so that the quantum gravity corrections to the background evolution are negligible.
For simplicity we disregard the contribution of baryons, which gives $c_s^2=1/3$~(this is a valid approximation at early times), and focus only on adiabatic perturbations, $(\delta p)_{\rm nad}=0$.
Thus, in Fourier space Eq.~\eqref{Eq:SecondOrderPDEPhi} boils down to 
\be\label{Eq:SecondOrderPDEPhiRDE}
\lf(1+\frac{8\alpha\kappa}{3A^2 \eta^4}\rg)\phi^{\prime\prime}+\lf(\frac{4}{\eta}-\frac{8\alpha\kappa}{3A^2\eta^5}\rg)\phi^{\prime}+\lf(\frac{k^2}{3}-\frac{40\alpha\kappa}{3A^2\eta^6}\rg)\phi=0~,
\ee
where $A=a(\eta_i)/\eta_i$, with $\eta_i$ some initial time.

If $\alpha<0$ the coefficient of $\phi^{\prime\prime}$ in Eq.~\eqref{Eq:SecondOrderPDEPhiRDE} becomes negative for
\be
\eta<\eta_{\star}\equiv\lf(\frac{8|\alpha|\kappa}{3A^2}\rg)^{1/4}~.
\ee
In order to avoid instabilities, we shall require $\eta_{\star}\ll \eta_{\rm reh}=1/{\cal H}_{\rm reh}$. This condition is automatically satisfied if one imposes the constraint~\eqref{Eq:AlphaConstraint}. Similarly, if $\alpha>0$ the coefficient of $\phi^{\prime}$ in Eq.~\eqref{Eq:SecondOrderPDEPhiRDE} may become negative; again, this can be prevented if \eqref{Eq:AlphaConstraint} is satisfied.

Lastly, we observe that with $\alpha>0$ also the coefficient of $\phi$ in Eq.~\eqref{Eq:SecondOrderPDEPhiRDE} may become negative at early times. To prevent this instability from affecting observable scales, we must impose a tighter constraint
\be\label{Eq:AlphaConstraint_FromPert1}
\alpha\ll \frac{M_{\rm Pl}^2}{40 H_{\rm reh}^2}\bar{k}^2 \eta_{\rm reh}^2=\frac{M_{\rm Pl}^2 H_o^2}{40 H_{\rm reh}^4} e^{2N}~,
\ee
where $\bar{k}\simeq a_o H_o$ is the comoving wave-number corresponding to the largest observable scales today and $N=\log(a_o/a_{\rm reh})$ is the number of e-folds since the time of reheating to the present time.
Expressing the Hubble rate at reheating as $H_{\rm reh}\simeq (g_{s}^{1/2}\pi/9.5) T_{\rm reh}^2/M_{\rm Pl}$ \cite{Lozanov:2019jxc}, and substituting the values $H_o^2\simeq 3.5\times10^{-121} M_{\rm Pl}^2$, $N\simeq60$, and $g_s\simeq 10^2$ for the effective number of relativistic species, we obtain
\be\label{Eq:AlphaConstraint_FromPert2}
\alpha\ll 10^{-72} \lf(\frac{M_{\rm Pl}}{T_{\rm reh}}\rg)^8~.
\ee
For instance, with a reheating temperature $T_{\rm reh}= \mathcal{O}(10^{15} {\rm GeV})$ we obtain the bound $\alpha\ll 10^{-45}$. However, the bound on $\alpha$ becomes less stringent with a lower reheating temperature, and  becomes very loose already for $T_{\rm reh}=\mathcal{O}(10^{9} {\rm GeV})$. Thus, if the reheating temperature is large enough, approaches to quantum gravity that predict $\alpha>0$ would be strongly disfavoured.

\section{Tensor Perturbations}\label{Sec:Tensor}
Similarly to our previous analysis of scalar perturbations in Sec.~\ref{Sec:Scalar}, here we study the modified dynamics of tensor perturbations.
We consider the perturbed line element
\be\label{Eq:FLRWmetricTensorPert}
\de s^2= a(\eta)^2 \Big(- \de \eta^2+ (\delta_{ij}+h_{ij}(\eta,x)) \de x^i\de x^j \Big) ~,
\ee
where $h^i_{\;i}=0$, $h_{ij}^{\;\;,j}=0$. Tensor perturbations are sourced by the tensorial part of the anisotropic stress, which contributes to the spatial components of the perturbed stress-energy tensor
\be\label{Eq:AnisotropicStress}
\delta T^i_{\pha j}=\pi^i_{\; j}~.
\ee
The anisotropic stress is traceless, $\pi^i_{\pha i}=0$. Substituting Eqs.~\eqref{Eq:FLRWmetricTensorPert} and \eqref{Eq:AnisotropicStress} in the field equations \eqref{Eq:FieldEquations}, we obtain
\be\label{Eq:TensorPertDynamics}
\Big(1+\frac{\alpha\kappa}{a^2}\lf({\cal H}^{\prime}-{\cal H}^2\rg) \Big)\lf(h_{ij}^{\prime\prime}+2{\cal H}h_{ij}^{\prime}-\Delta h_{ij}\rg)=2\kappa a^2 \pi_{ij}~.
\ee
Thus, the free propagation of gravitational waves is the same as in general relativity; however, their coupling to the anisotropic stress of matter is modified and depends on time. In particular, during radiation domination, Eq.~\eqref{Eq:TensorPertDynamics} boils down to
\be\label{Eq:TensorPer_RDE}
h_{ij}^{\prime\prime}+2{\cal H}h_{ij}^{\prime}-\Delta h_{ij}= \lf(1-2\alpha\kappa a^{-2} {\cal H}^2\rg)^{-1}2\kappa a^2 \pi_{ij}~.
\ee
However, the constraints \eqref{Eq:AlphaConstraint}, \eqref{Eq:AlphaConstraint_FromPert2} imply that the departures from the standard matter coupling in Eq.~\eqref{Eq:TensorPer_RDE} are too small to lead to observationally relevant effects.

\section{Discussion and Outlook}\label{Sec:Discussion}

In this paper we analyzed the cosmology of the ``quantum phenomenological gravitational dynamics'' model proposed in Ref.~\cite{Alonso-Serrano:2020dcz}, which includes corrections to the Raychaudhuri equations of order $\mathcal{O}(\dot{H}/M_{Pl}^2)$, see Eq.~\eqref{Eq:Hprime}. Such correction terms are accompanied by a dimensionless free parameter $\alpha$. Using dynamical system techniques, we showed that the model admits a family of solutions that are self-accelerating both at early and late times: they approach de Sitter in the future and admit a past attractor corresponding to an inflationary acceleration era.  Moreover, the model does not predict a bounce as a resolution of the initial cosmological singularity. This is in contrast with previous claims~\cite{Alonso-Serrano:2020dcz,Alonso-Serrano:2022nmb}, where the analysis of the background dynamics relied on a perturbative treatment of the corrections to general relativity; however, as shown above this assumption does not hold in the early universe. Therefore, this model provides an example where logarithmic corrections to the Bekenstein entropy law do not lead to a bouncing cosmology. 

Solutions that are viable at late times are represented by the green curves in Figs.~\ref{Fig:PhasePortrait_PosAlpha},~\ref{Fig:PhasePortrait_NegAlpha}. Both for positive and negative values of $\alpha$, these solutions asymptotically approach a de Sitter universe in the future, while they approach an inflationary attractor in the past that has different properties depending on the sign of $\alpha$. For $\alpha>0$ the initial singularity is not resolved, and the evolution on the past attractor is similar to power-law inflation if the dominant matter species has a non-relativistic equation of state. However, in the absence of a mechanism that explains why non-relativistic matter should be dominant before the radiation dominated era, this scenario is ruled out.\footnote{For instance, in some inflationary models, the equation of state is effectively $w=0$ at the end of inflation, when the inflaton is oscillating around its minimum~\cite{Lozanov:2019jxc}. In the case of inflation though there is a dynamical mechanism explaining the transition to $w=1/3$ during reheating.}
For $\alpha<0$ the initial singularity is also not resolved, even though the energy density approaches an upper bound given by Eq.~\eqref{Eq:MaximalRho}. 
However, the past attractor does not correspond to an epoch of slow-roll inflation, since the slow-roll conditions $\epsilon$, $|\eta|\ll1$ cannot be satisfied simultaneously; hence, the model cannot solve the horizon problem. We conclude that for any values of $\alpha$ the present model is unable to represent a valid alternative to the inflationary scenario. Therefore, even though the model admits solutions that are viable at late times, a completion in the past is required.
Nonetheless, it would be interesting to investigate whether these early-time issues may be addressed by including higher-order corrections in the low-energy expansion of the effective quantum gravity equations (inherited from additional corrections to the Bekenstein entropy) derived in Ref.~\cite{Alonso-Serrano:2020dcz}. In particular, future work should also clarify whether higher-order corrections may lead to a bounce, which is ruled out if only leading-order corrections are taken into account.

Summarizing, in its present formulation the model can at best be applied to the cosmological epochs following the time of reheating. In this regime, the condition that the corrections to the Raychaudhuri equations be sub-leading, so as not to spoil the predictions of standard cosmology, leads to mild bounds on the parameter $\alpha$, see Eq.~\eqref{Eq:AlphaConstraint}. This is due to the fact that the correction term in Eq.~\eqref{Eq:Hprime} is already suppressed by the Planck mass squared; moreover, the suppression becomes stronger as time increases.
Asking scalar perturbations to be unaffected by instabilities on observable scales leads to a much more stringent constraint in the case $\alpha>0$ if the reheating temperature is higher than ${\cal O}(10^9 {\rm GeV})$, see the discussion at the end of Section~\ref{Eq:ScalarPert_RDE}. Therefore, evidence for a reheating temperature above this threshold would strongly disfavour approaches to quantum gravity that predict a positive pre-factor for the logarithmic corrections to the Bekenstein entropy.
In the case $\alpha<0$ (which includes the case of Loop Quantum Gravity), the constraints are not as stringent and are given by Eq.~\eqref{Eq:AlphaConstraint}.
Finally, the dynamical equation for the propagation of tensor perturbations \eqref{Eq:TensorPertDynamics} entails a non-standard coupling to the anisotropic stress of matter, and thus belong to the general class of modified propagation equations studied in Ref.~\cite{Saltas:2014dha}; however, in this case the departures from general relativity are too strongly suppressed to be observationally relevant.
Interestingly, our results show that in the absence of anisotropic stress the dynamical equations for tensor perturbations are formally the same as in general relativity. This is in agreement with Ref.~\cite{Alonso-Serrano:2023xwr}, where however this was only taken as an assumption.

We conclude by noticing that, as pointed out in the Introduction, the model at hand can be regarded as a generalization of trace-free Einstein gravity. In both such theories, the conservation of energy-momentum does not follow from the field equations, but rather constitute a separate assumption. As argued in Ref.~\cite{Perez:2017krv}, quantum gravity effects may lead to a violation of energy-momentum conservation, which in turn provides a possible explanation of (dynamical) dark energy. In this paper we assumed, following Refs.~\cite{Alonso-Serrano:2020dcz,Alonso-Serrano:2022nmb}, that the energy-momentum is strictly conserved. The consequences of relaxing such an assumption in the model at hand deserve further investigation.

\section*{Acknowledgments}

We thank Ana Alonso-Serrano for discussions and Edward Wilson-Ewing for helpful comments on an earlier draft of our paper.
This work is supported by Ministero dell'Universit{\`a} e Ricerca (MUR) (Bando PRIN 2017, Codice Progetto: 20179ZF5K5\_006) and by INFN (Iniziativa specifica QUAGRAP and GeoSymQFT). Additionally, GG acknowledges financial support  by the Programme STAR Plus, funded by Federico II University and Compagnia di San Paolo. This work contributes to the European Union COST Action CA18108 {\it Quantum gravity phenomenology in the multi-messenger approach}.

\bibliography{quantumgravpheno}

\begin{thebibliography}{23}%
\makeatletter
\providecommand \@ifxundefined [1]{%
 \@ifx{#1\undefined}
}%
\providecommand \@ifnum [1]{%
 \ifnum #1\expandafter \@firstoftwo
 \else \expandafter \@secondoftwo
 \fi
}%
\providecommand \@ifx [1]{%
 \ifx #1\expandafter \@firstoftwo
 \else \expandafter \@secondoftwo
 \fi
}%
\providecommand \natexlab [1]{#1}%
\providecommand \enquote  [1]{``#1''}%
\providecommand \bibnamefont  [1]{#1}%
\providecommand \bibfnamefont [1]{#1}%
\providecommand \citenamefont [1]{#1}%
\providecommand \href@noop [0]{\@secondoftwo}%
\providecommand \href [0]{\begingroup \@sanitize@url \@href}%
\providecommand \@href[1]{\@@startlink{#1}\@@href}%
\providecommand \@@href[1]{\endgroup#1\@@endlink}%
\providecommand \@sanitize@url [0]{\catcode `\\12\catcode `\$12\catcode
  `\&12\catcode `\#12\catcode `\^12\catcode `\_12\catcode `\%12\relax}%
\providecommand \@@startlink[1]{}%
\providecommand \@@endlink[0]{}%
\providecommand \url  [0]{\begingroup\@sanitize@url \@url }%
\providecommand \@url [1]{\endgroup\@href {#1}{\urlprefix }}%
\providecommand \urlprefix  [0]{URL }%
\providecommand \Eprint [0]{\href }%
\providecommand \doibase [0]{http://dx.doi.org/}%
\providecommand \selectlanguage [0]{\@gobble}%
\providecommand \bibinfo  [0]{\@secondoftwo}%
\providecommand \bibfield  [0]{\@secondoftwo}%
\providecommand \translation [1]{[#1]}%
\providecommand \BibitemOpen [0]{}%
\providecommand \bibitemStop [0]{}%
\providecommand \bibitemNoStop [0]{.\EOS\space}%
\providecommand \EOS [0]{\spacefactor3000\relax}%
\providecommand \BibitemShut  [1]{\csname bibitem#1\endcsname}%
\let\auto@bib@innerbib\@empty
\bibitem [{\citenamefont {Kaul}\ and\ \citenamefont
  {Majumdar}(2000)}]{Kaul:2000kf}%
  \BibitemOpen
  \bibfield  {author} {\bibinfo {author} {\bibfnamefont {Romesh~K.}\
  \bibnamefont {Kaul}}\ and\ \bibinfo {author} {\bibfnamefont {Parthasarathi}\
  \bibnamefont {Majumdar}},\ }\bibfield  {title} {\enquote {\bibinfo {title}
  {{Logarithmic correction to the Bekenstein-Hawking entropy}},}\ }\href
  {\doibase 10.1103/PhysRevLett.84.5255} {\bibfield  {journal} {\bibinfo
  {journal} {Phys. Rev. Lett.}\ }\textbf {\bibinfo {volume} {84}},\ \bibinfo
  {pages} {5255--5257} (\bibinfo {year} {2000})},\ \Eprint
  {http://arxiv.org/abs/gr-qc/0002040} {arXiv:gr-qc/0002040} \BibitemShut
  {NoStop}%
\bibitem [{\citenamefont {Carlip}(2000)}]{Carlip:2000nv}%
  \BibitemOpen
  \bibfield  {author} {\bibinfo {author} {\bibfnamefont {Steven}\ \bibnamefont
  {Carlip}},\ }\bibfield  {title} {\enquote {\bibinfo {title} {{Logarithmic
  corrections to black hole entropy from the Cardy formula}},}\ }\href
  {\doibase 10.1088/0264-9381/17/20/302} {\bibfield  {journal} {\bibinfo
  {journal} {Class. Quant. Grav.}\ }\textbf {\bibinfo {volume} {17}},\ \bibinfo
  {pages} {4175--4186} (\bibinfo {year} {2000})},\ \Eprint
  {http://arxiv.org/abs/gr-qc/0005017} {arXiv:gr-qc/0005017} \BibitemShut
  {NoStop}%
\bibitem [{\citenamefont {Meissner}(2004)}]{Meissner:2004ju}%
  \BibitemOpen
  \bibfield  {author} {\bibinfo {author} {\bibfnamefont {Krzysztof~A.}\
  \bibnamefont {Meissner}},\ }\bibfield  {title} {\enquote {\bibinfo {title}
  {{Black hole entropy in loop quantum gravity}},}\ }\href {\doibase
  10.1088/0264-9381/21/22/015} {\bibfield  {journal} {\bibinfo  {journal}
  {Class. Quant. Grav.}\ }\textbf {\bibinfo {volume} {21}},\ \bibinfo {pages}
  {5245--5252} (\bibinfo {year} {2004})},\ \Eprint
  {http://arxiv.org/abs/gr-qc/0407052} {arXiv:gr-qc/0407052} \BibitemShut
  {NoStop}%
\bibitem [{\citenamefont {Sen}(2013)}]{Sen:2012dw}%
  \BibitemOpen
  \bibfield  {author} {\bibinfo {author} {\bibfnamefont {Ashoke}\ \bibnamefont
  {Sen}},\ }\bibfield  {title} {\enquote {\bibinfo {title} {{Logarithmic
  Corrections to Schwarzschild and Other Non-extremal Black Hole Entropy in
  Different Dimensions}},}\ }\href {\doibase 10.1007/JHEP04(2013)156}
  {\bibfield  {journal} {\bibinfo  {journal} {JHEP}\ }\textbf {\bibinfo
  {volume} {04}},\ \bibinfo {pages} {156} (\bibinfo {year} {2013})},\ \Eprint
  {http://arxiv.org/abs/1205.0971} {arXiv:1205.0971 [hep-th]} \BibitemShut
  {NoStop}%
\bibitem [{\citenamefont {Ashtekar}\ and\ \citenamefont
  {Singh}(2011)}]{Ashtekar:2011ni}%
  \BibitemOpen
  \bibfield  {author} {\bibinfo {author} {\bibfnamefont {Abhay}\ \bibnamefont
  {Ashtekar}}\ and\ \bibinfo {author} {\bibfnamefont {Parampreet}\ \bibnamefont
  {Singh}},\ }\bibfield  {title} {\enquote {\bibinfo {title} {{Loop Quantum
  Cosmology: A Status Report}},}\ }\href {\doibase
  10.1088/0264-9381/28/21/213001} {\bibfield  {journal} {\bibinfo  {journal}
  {Class. Quant. Grav.}\ }\textbf {\bibinfo {volume} {28}},\ \bibinfo {pages}
  {213001} (\bibinfo {year} {2011})},\ \Eprint {http://arxiv.org/abs/1108.0893}
  {arXiv:1108.0893 [gr-qc]} \BibitemShut {NoStop}%
\bibitem [{\citenamefont {Ashtekar}\ and\ \citenamefont
  {Gupt}(2015)}]{Ashtekar:2015iza}%
  \BibitemOpen
  \bibfield  {author} {\bibinfo {author} {\bibfnamefont {Abhay}\ \bibnamefont
  {Ashtekar}}\ and\ \bibinfo {author} {\bibfnamefont {Brajesh}\ \bibnamefont
  {Gupt}},\ }\bibfield  {title} {\enquote {\bibinfo {title} {{Generalized
  effective description of loop quantum cosmology}},}\ }\href {\doibase
  10.1103/PhysRevD.92.084060} {\bibfield  {journal} {\bibinfo  {journal} {Phys.
  Rev. D}\ }\textbf {\bibinfo {volume} {92}},\ \bibinfo {pages} {084060}
  (\bibinfo {year} {2015})},\ \Eprint {http://arxiv.org/abs/1509.08899}
  {arXiv:1509.08899 [gr-qc]} \BibitemShut {NoStop}%
\bibitem [{\citenamefont {Alonso-Serrano}\ and\ \citenamefont
  {Li\v{s}ka}(2020)}]{Alonso-Serrano:2020dcz}%
  \BibitemOpen
  \bibfield  {author} {\bibinfo {author} {\bibfnamefont {Ana}\ \bibnamefont
  {Alonso-Serrano}}\ and\ \bibinfo {author} {\bibfnamefont {Marek}\
  \bibnamefont {Li\v{s}ka}},\ }\bibfield  {title} {\enquote {\bibinfo {title}
  {{Quantum phenomenological gravitational dynamics: A general view from
  thermodynamics of spacetime}},}\ }\href {\doibase 10.1007/JHEP12(2020)196}
  {\bibfield  {journal} {\bibinfo  {journal} {JHEP}\ }\textbf {\bibinfo
  {volume} {12}},\ \bibinfo {pages} {196} (\bibinfo {year} {2020})},\ \Eprint
  {http://arxiv.org/abs/2009.03826} {arXiv:2009.03826 [gr-qc]} \BibitemShut
  {NoStop}%
\bibitem [{\citenamefont {Jacobson}(1995)}]{Jacobson:1995ab}%
  \BibitemOpen
  \bibfield  {author} {\bibinfo {author} {\bibfnamefont {Ted}\ \bibnamefont
  {Jacobson}},\ }\bibfield  {title} {\enquote {\bibinfo {title}
  {{Thermodynamics of space-time: The Einstein equation of state}},}\ }\href
  {\doibase 10.1103/PhysRevLett.75.1260} {\bibfield  {journal} {\bibinfo
  {journal} {Phys. Rev. Lett.}\ }\textbf {\bibinfo {volume} {75}},\ \bibinfo
  {pages} {1260--1263} (\bibinfo {year} {1995})},\ \Eprint
  {http://arxiv.org/abs/gr-qc/9504004} {arXiv:gr-qc/9504004} \BibitemShut
  {NoStop}%
\bibitem [{\citenamefont {Alonso-Serrano}\ \emph
  {et~al.}(2023{\natexlab{a}})\citenamefont {Alonso-Serrano}, \citenamefont
  {Li\v{s}ka},\ and\ \citenamefont
  {Vicente-Becerril}}]{Alonso-Serrano:2022nmb}%
  \BibitemOpen
  \bibfield  {author} {\bibinfo {author} {\bibfnamefont {Ana}\ \bibnamefont
  {Alonso-Serrano}}, \bibinfo {author} {\bibfnamefont {Marek}\ \bibnamefont
  {Li\v{s}ka}}, \ and\ \bibinfo {author} {\bibfnamefont {Antonio}\ \bibnamefont
  {Vicente-Becerril}},\ }\bibfield  {title} {\enquote {\bibinfo {title}
  {{Friedmann equations and cosmic bounce in a modified cosmological
  scenario}},}\ }\href {\doibase 10.1016/j.physletb.2023.137827} {\bibfield
  {journal} {\bibinfo  {journal} {Phys. Lett. B}\ }\textbf {\bibinfo {volume}
  {839}},\ \bibinfo {pages} {137827} (\bibinfo {year} {2023}{\natexlab{a}})},\
  \Eprint {http://arxiv.org/abs/2212.10928} {arXiv:2212.10928 [gr-qc]}
  \BibitemShut {NoStop}%
\bibitem [{\citenamefont {Brandenberger}\ and\ \citenamefont
  {Peter}(2017)}]{Brandenberger:2016vhg}%
  \BibitemOpen
  \bibfield  {author} {\bibinfo {author} {\bibfnamefont {Robert}\ \bibnamefont
  {Brandenberger}}\ and\ \bibinfo {author} {\bibfnamefont {Patrick}\
  \bibnamefont {Peter}},\ }\bibfield  {title} {\enquote {\bibinfo {title}
  {{Bouncing Cosmologies: Progress and Problems}},}\ }\href {\doibase
  10.1007/s10701-016-0057-0} {\bibfield  {journal} {\bibinfo  {journal} {Found.
  Phys.}\ }\textbf {\bibinfo {volume} {47}},\ \bibinfo {pages} {797--850}
  (\bibinfo {year} {2017})},\ \Eprint {http://arxiv.org/abs/1603.05834}
  {arXiv:1603.05834 [hep-th]} \BibitemShut {NoStop}%
\bibitem [{\citenamefont {Ellis}\ \emph {et~al.}(2011)\citenamefont {Ellis},
  \citenamefont {van Elst}, \citenamefont {Murugan},\ and\ \citenamefont
  {Uzan}}]{Ellis:2010uc}%
  \BibitemOpen
  \bibfield  {author} {\bibinfo {author} {\bibfnamefont {George F.~R.}\
  \bibnamefont {Ellis}}, \bibinfo {author} {\bibfnamefont {Henk}\ \bibnamefont
  {van Elst}}, \bibinfo {author} {\bibfnamefont {Jeff}\ \bibnamefont
  {Murugan}}, \ and\ \bibinfo {author} {\bibfnamefont {Jean-Philippe}\
  \bibnamefont {Uzan}},\ }\bibfield  {title} {\enquote {\bibinfo {title} {{On
  the Trace-Free Einstein Equations as a Viable Alternative to General
  Relativity}},}\ }\href {\doibase 10.1088/0264-9381/28/22/225007} {\bibfield
  {journal} {\bibinfo  {journal} {Class. Quant. Grav.}\ }\textbf {\bibinfo
  {volume} {28}},\ \bibinfo {pages} {225007} (\bibinfo {year} {2011})},\
  \Eprint {http://arxiv.org/abs/1008.1196} {arXiv:1008.1196 [gr-qc]}
  \BibitemShut {NoStop}%
\bibitem [{\citenamefont {Henneaux}\ and\ \citenamefont
  {Teitelboim}(1989)}]{HENNEAUX1989195}%
  \BibitemOpen
  \bibfield  {author} {\bibinfo {author} {\bibfnamefont {Marc}\ \bibnamefont
  {Henneaux}}\ and\ \bibinfo {author} {\bibfnamefont {Claudio}\ \bibnamefont
  {Teitelboim}},\ }\bibfield  {title} {\enquote {\bibinfo {title} {The
  cosmological constant and general covariance},}\ }\href {\doibase
  https://doi.org/10.1016/0370-2693(89)91251-3} {\bibfield  {journal} {\bibinfo
   {journal} {Physics Letters B}\ }\textbf {\bibinfo {volume} {222}},\ \bibinfo
  {pages} {195--199} (\bibinfo {year} {1989})}\BibitemShut {NoStop}%
\bibitem [{\citenamefont {Kucha{\v r}}(1991)}]{Kuchar:1991aa}%
  \BibitemOpen
  \bibfield  {author} {\bibinfo {author} {\bibfnamefont {Karel~V.}\
  \bibnamefont {Kucha{\v r}}},\ }\bibfield  {title} {\enquote {\bibinfo {title}
  {Does an unspecified cosmological constant solve the problem of time in
  quantum gravity?}}\ }\href {\doibase 10.1103/PhysRevD.43.3332} {\bibfield
  {journal} {\bibinfo  {journal} {Physical Review D}\ }\textbf {\bibinfo
  {volume} {43}},\ \bibinfo {pages} {3332--3344} (\bibinfo {year}
  {1991})}\BibitemShut {NoStop}%
\bibitem [{\citenamefont {Percacci}(2018)}]{Percacci:2017fsy}%
  \BibitemOpen
  \bibfield  {author} {\bibinfo {author} {\bibfnamefont {R.}~\bibnamefont
  {Percacci}},\ }\bibfield  {title} {\enquote {\bibinfo {title} {{Unimodular
  quantum gravity and the cosmological constant}},}\ }\href {\doibase
  10.1007/s10701-018-0189-5} {\bibfield  {journal} {\bibinfo  {journal} {Found.
  Phys.}\ }\textbf {\bibinfo {volume} {48}},\ \bibinfo {pages} {1364--1379}
  (\bibinfo {year} {2018})},\ \Eprint {http://arxiv.org/abs/1712.09903}
  {arXiv:1712.09903 [gr-qc]} \BibitemShut {NoStop}%
\bibitem [{\citenamefont {Leon}(2022)}]{Leon:2022kwn}%
  \BibitemOpen
  \bibfield  {author} {\bibinfo {author} {\bibfnamefont {Gabriel}\ \bibnamefont
  {Leon}},\ }\bibfield  {title} {\enquote {\bibinfo {title} {{Inflation and the
  cosmological (not-so) constant in unimodular gravity}},}\ }\href {\doibase
  10.1088/1361-6382/ac52bc} {\bibfield  {journal} {\bibinfo  {journal} {Class.
  Quant. Grav.}\ }\textbf {\bibinfo {volume} {39}},\ \bibinfo {pages} {075008}
  (\bibinfo {year} {2022})},\ \Eprint {http://arxiv.org/abs/2202.04029}
  {arXiv:2202.04029 [gr-qc]} \BibitemShut {NoStop}%
\bibitem [{\citenamefont {Josset}\ \emph {et~al.}(2017)\citenamefont {Josset},
  \citenamefont {Perez},\ and\ \citenamefont {Sudarsky}}]{Josset:2016vrq}%
  \BibitemOpen
  \bibfield  {author} {\bibinfo {author} {\bibfnamefont {Thibaut}\ \bibnamefont
  {Josset}}, \bibinfo {author} {\bibfnamefont {Alejandro}\ \bibnamefont
  {Perez}}, \ and\ \bibinfo {author} {\bibfnamefont {Daniel}\ \bibnamefont
  {Sudarsky}},\ }\bibfield  {title} {\enquote {\bibinfo {title} {{Dark Energy
  from Violation of Energy Conservation}},}\ }\href {\doibase
  10.1103/PhysRevLett.118.021102} {\bibfield  {journal} {\bibinfo  {journal}
  {Phys. Rev. Lett.}\ }\textbf {\bibinfo {volume} {118}},\ \bibinfo {pages}
  {021102} (\bibinfo {year} {2017})},\ \Eprint
  {http://arxiv.org/abs/1604.04183} {arXiv:1604.04183 [gr-qc]} \BibitemShut
  {NoStop}%
\bibitem [{\citenamefont {Perez}\ \emph {et~al.}(2021)\citenamefont {Perez},
  \citenamefont {Sudarsky},\ and\ \citenamefont
  {Wilson-Ewing}}]{Perez:2020cwa}%
  \BibitemOpen
  \bibfield  {author} {\bibinfo {author} {\bibfnamefont {Alejandro}\
  \bibnamefont {Perez}}, \bibinfo {author} {\bibfnamefont {Daniel}\
  \bibnamefont {Sudarsky}}, \ and\ \bibinfo {author} {\bibfnamefont {Edward}\
  \bibnamefont {Wilson-Ewing}},\ }\bibfield  {title} {\enquote {\bibinfo
  {title} {{Resolving the $H_0$ tension with diffusion}},}\ }\href {\doibase
  10.1007/s10714-020-02781-0} {\bibfield  {journal} {\bibinfo  {journal} {Gen.
  Rel. Grav.}\ }\textbf {\bibinfo {volume} {53}},\ \bibinfo {pages} {7}
  (\bibinfo {year} {2021})},\ \Eprint {http://arxiv.org/abs/2001.07536}
  {arXiv:2001.07536 [astro-ph.CO]} \BibitemShut {NoStop}%
\bibitem [{\citenamefont {de~Cesare}\ and\ \citenamefont
  {Wilson-Ewing}(2022)}]{deCesare:2021wmk}%
  \BibitemOpen
  \bibfield  {author} {\bibinfo {author} {\bibfnamefont {Marco}\ \bibnamefont
  {de~Cesare}}\ and\ \bibinfo {author} {\bibfnamefont {Edward}\ \bibnamefont
  {Wilson-Ewing}},\ }\bibfield  {title} {\enquote {\bibinfo {title}
  {{Interacting dark sector from the trace-free Einstein equations:
  Cosmological perturbations with no instability}},}\ }\href {\doibase
  10.1103/PhysRevD.106.023527} {\bibfield  {journal} {\bibinfo  {journal}
  {Phys. Rev. D}\ }\textbf {\bibinfo {volume} {106}},\ \bibinfo {pages}
  {023527} (\bibinfo {year} {2022})},\ \Eprint
  {http://arxiv.org/abs/2112.12701} {arXiv:2112.12701 [gr-qc]} \BibitemShut
  {NoStop}%
\bibitem [{\citenamefont {Perez}\ and\ \citenamefont
  {Sudarsky}(2019)}]{Perez:2017krv}%
  \BibitemOpen
  \bibfield  {author} {\bibinfo {author} {\bibfnamefont {Alejandro}\
  \bibnamefont {Perez}}\ and\ \bibinfo {author} {\bibfnamefont {Daniel}\
  \bibnamefont {Sudarsky}},\ }\bibfield  {title} {\enquote {\bibinfo {title}
  {{Dark energy from quantum gravity discreteness}},}\ }\href {\doibase
  10.1103/PhysRevLett.122.221302} {\bibfield  {journal} {\bibinfo  {journal}
  {Phys. Rev. Lett.}\ }\textbf {\bibinfo {volume} {122}},\ \bibinfo {pages}
  {221302} (\bibinfo {year} {2019})},\ \Eprint
  {http://arxiv.org/abs/1711.05183} {arXiv:1711.05183 [gr-qc]} \BibitemShut
  {NoStop}%
\bibitem [{\citenamefont {Wainwright}\ and\ \citenamefont
  {Ellis}(1997)}]{wainwright_ellis_1997}%
  \BibitemOpen
  \bibinfo {editor} {\bibfnamefont {J.}~\bibnamefont {Wainwright}}\ and\
  \bibinfo {editor} {\bibfnamefont {G.~F.~R.}\ \bibnamefont {Ellis}},\ eds.,\
  \href {\doibase 10.1017/CBO9780511524660} {\emph {\bibinfo {title} {Dynamical
  Systems in Cosmology}}}\ (\bibinfo  {publisher} {Cambridge University
  Press},\ \bibinfo {year} {1997})\BibitemShut {NoStop}%
\bibitem [{\citenamefont {Lozanov}(2019)}]{Lozanov:2019jxc}%
  \BibitemOpen
  \bibfield  {author} {\bibinfo {author} {\bibfnamefont {Kaloian~D.}\
  \bibnamefont {Lozanov}},\ }\bibfield  {title} {\enquote {\bibinfo {title}
  {{Lectures on Reheating after Inflation}},}\ }\href@noop {} {\  (\bibinfo
  {year} {2019})},\ \Eprint {http://arxiv.org/abs/1907.04402} {arXiv:1907.04402
  [astro-ph.CO]} \BibitemShut {NoStop}%
\bibitem [{\citenamefont {Saltas}\ \emph {et~al.}(2014)\citenamefont {Saltas},
  \citenamefont {Sawicki}, \citenamefont {Amendola},\ and\ \citenamefont
  {Kunz}}]{Saltas:2014dha}%
  \BibitemOpen
  \bibfield  {author} {\bibinfo {author} {\bibfnamefont {Ippocratis~D.}\
  \bibnamefont {Saltas}}, \bibinfo {author} {\bibfnamefont {Ignacy}\
  \bibnamefont {Sawicki}}, \bibinfo {author} {\bibfnamefont {Luca}\
  \bibnamefont {Amendola}}, \ and\ \bibinfo {author} {\bibfnamefont {Martin}\
  \bibnamefont {Kunz}},\ }\bibfield  {title} {\enquote {\bibinfo {title}
  {{Anisotropic Stress as a Signature of Nonstandard Propagation of
  Gravitational Waves}},}\ }\href {\doibase 10.1103/PhysRevLett.113.191101}
  {\bibfield  {journal} {\bibinfo  {journal} {Phys. Rev. Lett.}\ }\textbf
  {\bibinfo {volume} {113}},\ \bibinfo {pages} {191101} (\bibinfo {year}
  {2014})},\ \Eprint {http://arxiv.org/abs/1406.7139} {arXiv:1406.7139
  [astro-ph.CO]} \BibitemShut {NoStop}%
\bibitem [{\citenamefont {Alonso-Serrano}\ \emph
  {et~al.}(2023{\natexlab{b}})\citenamefont {Alonso-Serrano}, \citenamefont
  {Mena~Marugan},\ and\ \citenamefont
  {Vicente-Becerril}}]{Alonso-Serrano:2023xwr}%
  \BibitemOpen
  \bibfield  {author} {\bibinfo {author} {\bibfnamefont {Ana}\ \bibnamefont
  {Alonso-Serrano}}, \bibinfo {author} {\bibfnamefont {Guillermo~A.}\
  \bibnamefont {Mena~Marugan}}, \ and\ \bibinfo {author} {\bibfnamefont
  {Antonio}\ \bibnamefont {Vicente-Becerril}},\ }\bibfield  {title} {\enquote
  {\bibinfo {title} {{Primordial Power Spectrum in Modified Cosmology: From
  Thermodynamics of Spacetime to Loop Quantum Cosmology}},}\ }\href@noop {} {\
  (\bibinfo {year} {2023}{\natexlab{b}})},\ \Eprint
  {http://arxiv.org/abs/2307.06813} {arXiv:2307.06813 [gr-qc]} \BibitemShut
  {NoStop}%
\end{thebibliography}%

\end{document}